\documentclass{article}

\usepackage{RR}
\usepackage{graphicx}
\usepackage{hyperref}
\graphicspath{{figures/}}
\RRtitle{Utiliser des simulateurs de réseaux existants pour des réseaux de capteurs sans fils, efficaces en énergie et auto-organisés}
\RRetitle{Using Existing Network Simulators for Power-Aware Self-Organizing Wireless Sensor Network Protocols}
\titlehead{Comparing Wireless Sensor Networks Simulators}

\RRauthor{
   Thomas Watteyne\thanks[sfn]{CITI Laboratory, INSA Lyon, France. thomas.watteyne@insa-lyon.fr}\thanks[sfn]{France Telecom R\&D, Grenoble, France. thomas.watteyne@orange-ft.com}
}
\authorhead{Watteyne}

\RRdate{September 2006}

\RRresume{Dans ce document, nous comparons trois plateformes de simulation existantes (OPNET Modeler, Network Simulator 2, Georgia Tech Sensor Network Simulator). Notre étude comparative porte essentiellement sur la facilité d'utilisation, le passage à l'échelle, la facilité d'implémenter un modèle de consommation énergétique et la finesse du modèle de propagation radio utilisé. Les conclusions de cette étude nous aident à choisir une plateforme de simulation pour évaluer des protocoles d'auto-organisation efficaces en énergie pour réseaux de capteurs.}

\RRmotcle{Réseaux de Capteurs sans Fils, Simulateurs.}

\RRabstract{In this document, we compare three existing simulation platforms (OPNET Modeler, Network Simulator 2, Georgia Tech Sensor Network Simulator). Our comparative study focuses on ease of use, scalability, ease of implementing power consumption model and physical layer modeling accuracy, mainly. Conclusions of this study are presented, and will help us decide which simulating environment to use for evaluating power-aware self-organizing sensor networks protocols.}

\RRkeyword{Wireless Sensor Networks, Simulators.}

\RRprojet{ARES}
\RRtheme{\THCom}
\URRhoneAlpes

\begin{document}

\makeRR


\section{Introduction and constraints}
\label{section:Introduction and constraints}

We are interested in evaluating power-aware self-organizing protocols for Wireless Sensor Networks (WSNs). Although analytic methods are useful in
many situations, the complexity of modern networks combined with the inability to apply simplifying assumptions in many analysis problems (e.g., it is well known that Markovian traffic assumptions are often inappropriate and can lead to misleading results) limit the applicability of purely analytic approaches \cite{fujimoto03how}.

Nevertheless, as has been shown in \cite{cavin02accuracy, haq05simulation, shnayder04simulating}, simulators do not always reflect the performances of the processes they model. Results are mainly corrupted by inaccuracies in the physical model (interferences, error models, power consumption, \ldots). This does not mean simulations should not be used, they only stress out that simulation results only give a rough idea of what is really going on. What's more, as simulations are performed on discrete topologies with discrete scenarios, simulation can never be used in a validation approach.

Several simulation platforms exist. The aim of this paper is to compare three of them (OPNET Modeler \cite{siteOPNET}, Network Simulator 2 \cite{siteNs2} and GTSNetS \cite{siteGtnets}), in order to pick out the one that best fulfills our needs.

WSN are very constrained, so are WSN simulators. As WSN communicate wirelessly, it is essential that the simulator embeds an accurate \textbf{propagation model}. What's more, depending on the final application, WSNs can be composed of tens of thousands nodes, which means \textbf{simulator scalability} is important too. As WSNs are very application specific, no standard solution can be used, or at least it should be possible to tune those solutions. \textbf{Extensibility} of the model is therefore essential. We are interesting in the power consumption of self-organizing protocols, so a proper \textbf{power consumption model} should be implemented in the simulator. Finally, the simulator should be \textbf{easy} to use and to learn, and its \textbf{cost} should be kept as low as possible.


\section{OPNET Modeler}

OPNET Modeler \cite{siteOPNET} is a simulation environment developed and maintained by OPNET Technologies, Inc. (NASDAQ: OPNT). Its aim is to enable users to evaluate how networking equipment, communications technologies, systems, and protocols perform under simulated network conditions. It is therefore mainly used by Service providers (AT\&T, BT, Deutsche Telekom, France Telecom, NTT DoCoMo \ldots), Entreprises (DaimlerChrysler, IBM Global Services, Microsoft, Oracle, Schneider Electric \ldots), Network-Equipment Manufacturers (Alcatel, Cisco Systems, Intel Corporation, Motorola, Nokia \ldots) and Government \& Defense (DARPA, Federal Aviation Administration, Federal Bureau of Investigation, NASA, NATO \ldots). It is a commercial product, and a license is far from free. Use of a educational license in cooperation with an industrial partner is prohibited.

The OPNET Modeler simulation environment can be divided in three parts: the modeling environment, the simulation engine and the result collector.


\subsection{The modeling environment}

Modeling is greatly simplified by building the model hierarchically. As communication has been cut into several OSI layers \cite{tanenbaum02computer}, a model built using OPNET Modeler is composed of several interchangeable layers. It is for example very easy to modify the propagation model being used by modifying the \emph{closure model} parameter used in the emitting layer.

The highest layer is the \emph{scenario model}. A scenario is composed of a node topology, and global simulation attributes.

The \emph{node level model} is described by a graphical representation of its processes. Those processes communicate using interrupts. Node level models also have attributes, such as geographic node position. An interface is present to let the node level model communicate with the higher model (the scenario). This way, a node model attribute can be initialized by the scenario when entering run-time, and this depending on which scenario is used.

\begin{figure}
\centering
\includegraphics[width=0.60\textwidth]{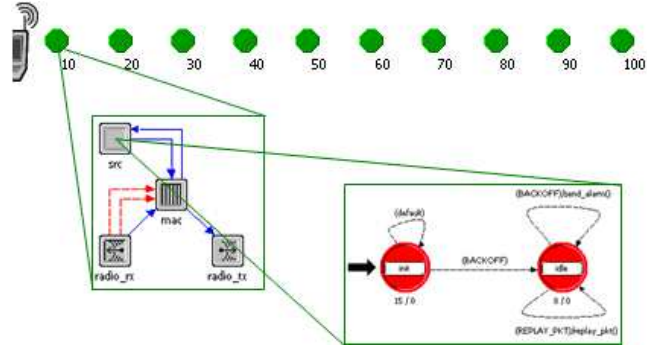}
\caption{Modeling in OPNET is done using layers.}
\label{figure:opnet_models}
\end{figure}

As in the node model, an interface and attributes are present in the \emph{process level model}. The process is described graphically by a states interconnected by conditional transitions. Time can only flow in states. A transition is fired only when its given conditions are satisfied. What's more, it is possible to perform computation on a given variable while firing a transition, when entering a given state, or when leaving it. Three major code blocks as defined: \emph{SV} (State Variable) which described the process variables, HB (Header Block) which describes the conditions to fire the transitions, and FB (Function Block) which contains all the functions' bodies. FB is written in C programming language, with an added OPNET Modeler function library. Choosing C as a programming language gives access to all classical C functions (for manipulating files, generating pseudo-random numbers \ldots). The layered model structure is depicted in Figure \ref{figure:opnet_models}.

\subsection{The simulation engine}

OPNET Modeler is a discrete event simulator. The complete system's behavior is represented internally by a sequence of events. An given execution time is attached to each event. During execution, a list of upcoming events is thus attached to each entity. When this execution time is reached, the processing attached to the events is performed, and the event is removed from the list. A new event can be scheduled (\textit{i.e.} attached to the  list) directly by using OPNET functions, or indirectly for example when a process receives an interrupt. Even if the notion of discrete event simulation is internal to the simulator, it is important for the user as she or he can directly interact using the appropriate functions on the list of events (adding, removing, filtering a given list of events).


\subsection{The result collector}

The OPNET Modeler environment also includes a result collector. This can calculate output statistics, and plot graphs while the simulation is running. Using specific functions, probes are placed at the appropriate locations in the code so as to measure a given value. These probes feed the result collector using vectorial or scalar result files. Afterward, OPNET Modeler can plot those files. Despite this integrated tool, we have not used it for its lack of flexibility. Indeed, you are bound to the functions OPNET Modeler proposes, and it is for example very unnatural to perform complex post-simulation calculations. We have therefore decided to collect output data using standard C functions, and writing data in output ASCII files. Post-processing and plotting can then be easily performed by software such as gnuplot \cite{siteGnuplot}.


\subsection{The physical layer model}

As we have seen with \cite{cavin02accuracy}, there is a great impact of the physical model on overall results in WSNs. So a good model of the radio interface is essential. Radio interface is represented in OPNET Modeler by a pipeline: the physical models (antenna gain, delays \ldots) are applied sequentially on the 11 stage long pipeline. Each of these stages is configurable, and this for emitting and receiving radio entities.

\begin{figure}
\centering
\includegraphics[width=0.60\textwidth]{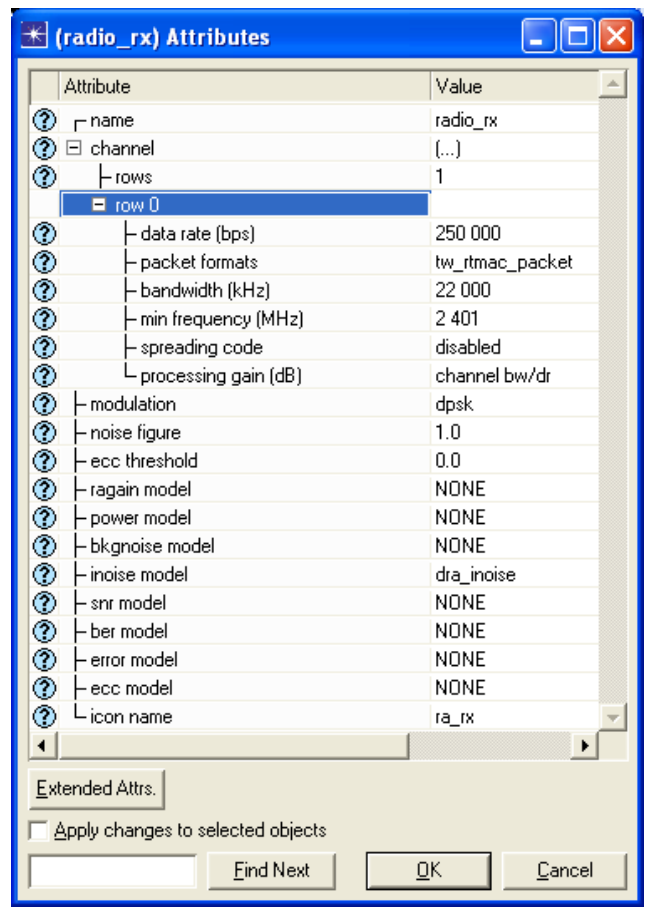}
\caption{Many parameters can be configured for the reception side of the radio.}
\label{figure:opnet_radio_parameters}
\end{figure}

The configurable pipeline-based radio model (see Figure \ref{figure:opnet_radio_parameters}) makes it possible to suppress uninteresting phenomena (or phenomena which are not taken into account yet in a given communication protocol). For simple Unit Disk Graph transmission model, it is possible to turn off power control (\textit{tagain}, \textit{ragain}, \textit{ragain model}, \textit{power model}), errors (\textit{error model}, \textit{ecc model}), and noise (\textit{bkgnoise model}, \textit{snr model}, \textit{ber model}). The \textit{closure model} parameter defines the actual propagation model, with \textit{mil\_closure\_dist} being the unit disk graph model.

\subsection{Conclusions on OPNET Modeler}

Thanks to our CITI's educational license, we have worked with OPNET Modeler. Even if it has been primarily designed for network planning, it can be used easily for wireless ad-hoc protocols providing its wireless extension is installed. The main advantage of this simulation platform is the ease of use. Indeed OPNET provides extensive documentation, and a comprehensible graphical modeling interface. Even with a mixture between pure graphical (scenario and node model) and textual code-based modeling (process model), modeling stays straightforward. Nevertheless, there are countless parameters and sometimes the line between model interfaces and global/local parameters gets fuzzy, yielding to moments of great confusion.

As for scalability according to \cite{fujimoto03how} OPNET Modeler is able to model networks of several hundreds to several thousand nodes. Of course scalability depends on numerous parameters such as the model refinement, available memory, processing power, available time etc. We will more specifically address scalability in Section \ref{section:Georgia Tech Sensor Network Simulator}.

Despite the fact the propagation model seems correct, the lack of scalability and the price of a exploitation license make us reject OPNET Modeler as a simulation environment.


\section{Network Simulator 2}

Ns \cite{siteNs2} is a discrete event simulator targeted at networking research. Ns provides substantial support for simulation of TCP, routing, and multicast protocols over wired and wireless (local and satellite) networks. 

Ns began as a variant of the REAL network simulator in 1989 and has evolved substantially over the past few years. In 1995 ns development was supported by DARPA through the VINT project at LBL, Xerox PARC, UCB, and USC/ISI. Currently ns development is support through DARPA with SAMAN and through NSF with CONSER, both in collaboration with other researchers including ACIRI. Ns has always included substantial contributions from other researchers, including wireless code from the UCB Daedelus and CMU Monarch projects and Sun Microsystems. For documentation, see \cite{fall06ns}. 

Ns-2 is free and open-source software. It is extensively used in educational/research institutions. Contributors are very numerous, so included packages come from very different people. The main advantage is that ns-2 includes cutting-edge modeled protocols. The main drawback is the complexity of resulting group of packages and the existence of untracked bugs. The same happens for documentation: despite an effort for centralizing information (for example in the new ns-2 manual \cite{fall06ns}), information/documentation is still disseminated on numerous Web sites. It is generally advies to seek information on the ns-2 dedicated forums, rather than written down manuals.

Ns-2 has no graphical interface, nor has it a layer modeling approach as OPNET Modeler for example. This lack of ease of use can scare away potential users. ns-2 uses two complementary object-oriented programming languages, a idea known as split-language programming: C++ -- a compiled languages -- is used for long and complex run-time phases, while OTcL -- a interpreted language -- is slower but supports frequent parameter changes.


\subsection{Modeling the radio interface}


\subsubsection{Radio propagation models}

Ns was first dedicated to simulating wired networks. Wi\-reless support was made possible by porting the CMU's Monarch group's mobility extension to ns. A node will only receive a signal if it is received above a certain power threshold. Reception power is calculated thanks to a propagation model. Three standard propagation models are provided with ns-2. The first one is the Friss free space model (Eq. \ref{eq:free_space}), which is not really realistic in everyday-use. The two-way ground effect model considers the radio wave travels using two paths: the direct line of sight path and a path which bounces off the ground (Equation \ref{eq:two_way}). A more realistic model would include the random fading effect, due to multipath propagation effects. We will not explicit those equations here.

\begin{equation}
P_{r}(d)=\frac{P_{t}G_{t}G_{r}\lambda^{2}}{(4\pi)^{2}d^{2}L}
\label{eq:free_space}
\end{equation}

\begin{equation}
P_{r}(d)=\frac{P_{t}G_{t}G_{r}h_{t}^{2}h_{r}^{2}}{d^{4}L}
\label{eq:two_way}
\end{equation}


\subsubsection{Energy models}

Basic energy models are very simple in ns-2. The node has a bucket of energy. When a message is sent/received, some energy is taken out of the bucket. When it empties, no more messages can be sent nor received. This energy model is of course much to scarce for Wireless Sensor Network applications for example, as it only takes into account the packet-based model, and not the time-based model, nor the physical model (see Section \ref{section:Introduction and constraints}).


\subsection{Add-ons for ns-2}

Because of the open-source nature of ns-2 contributions are added daily. We here give the example of visualization tools.

\begin{figure}
\centering
\includegraphics[width=0.60\textwidth]{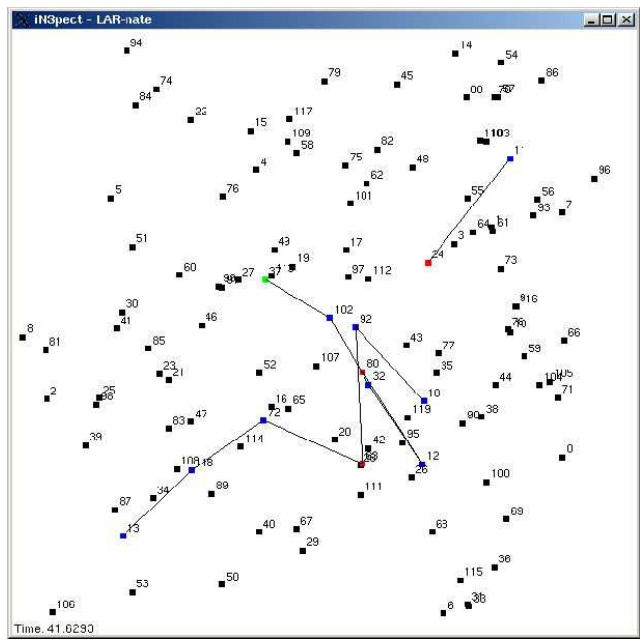}
\caption{iNSpect given you a fair idea of wait is going on in your network [drawn from \cite{kurkowski05visualization}].}
\label{figure:inspect}
\end{figure}

The "official" visualization tool for ns-2 is called Nam \cite{siteNam}. Whereas is has been extensively used for research on wired networks, it has not been extended for wireless visualization. Kurkowski et al. has recently proposed a visualization tool called iNSpect \cite{kurkowski05visualization} (see Figure \ref{figure:inspect}). The authors argue that this visualization tool can be used for (1) validating the accuracy of a mobility model's output and/or the node topology files used to drive the simulation; (2) validate the new versions of the NS-2 simulator itself; and (3) analyze the results of NS-2 simulations.

\subsection{Conclusions on ns-2}

As for scalability, according to \cite{riley03large}, on single processor machine, ns-2 can comfortably simulate networks of about 1000 nodes with popular routing protocols \cite{riley03large}. We will describe scalability more in detail in \ref{section:Georgia Tech Sensor Network Simulator}.

Ns-2 looks like a very good candidate for evaluating power-aware self-organizing wireless sensor networks protocols. It is free and open-source, which means the community of users is continuously updating the simulator with the latest protocols. Nevertheless, it is hard to use. What's more, for Wireless Sensor Network evaluation, ns-2 lacks scalability and satisfactory energy models.


\section{Georgia Tech Sensor Network Simulator}
\label{section:Georgia Tech Sensor Network Simulator}

The \textit{Georgia Tech Network Simulator} (GTNetS) is currently developped by Pr. George Riley and his students. It is freely available online \cite{siteGtnets} and was first presented in 2003 \cite{riley03georgia}. Before developping this new simulation environment, Pr. Riley was working on scalability issues in simulating networks. Scalability can be a main issue when trying to simulate wired peer-to-peer systems or wireless sensor networks composed of tens of thousand nodes.

\begin{figure}
\centering
\includegraphics[width=0.60\textwidth]{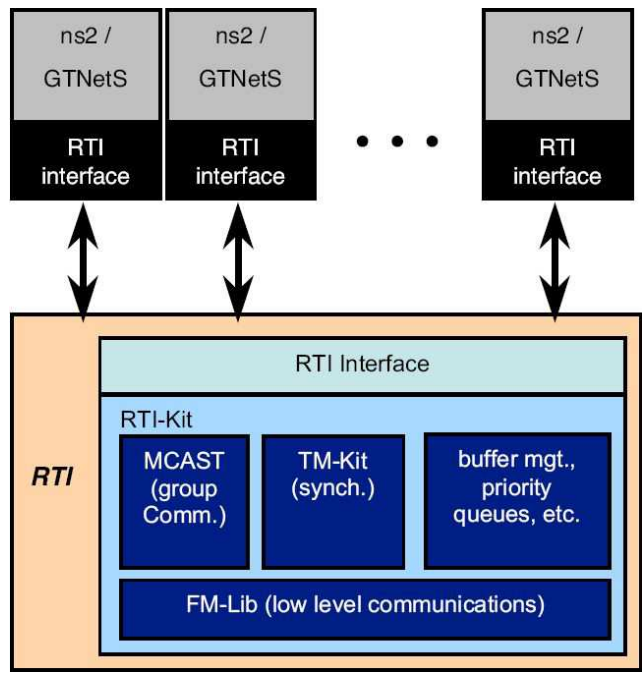}
\caption{Building a architecture to link ns-2 simulators together is not an easy task [drawn from \cite{fujimoto03how}].}
\label{figure:rti}
\end{figure}

Scalability is limited mainly by available memory and computation time. Another important factor is how fine-grained simulation models are built. As single processor PC-like computer are very limited, it was interesting to study the feasibility of parallel simulation. A first attempt was the \textit{Parallel/Distributed Network Simulator} in which a runtime infrastructure was developed in order to make several ns-2 simulators running on separate machine work together on the same simulation task (see Fig. \ref{figure:rti}). But the resulting architecture was not "natural", as ns-2 is not intended to be used in a distributed fashion, so GTNetS was developed. Its main goals are:

\begin{itemize}
   \item distributed simulation and scalability
   \item ease of use and extensibility
   \item support for popular protocols
\end{itemize}

\subsection{distributed simulation and scalability}

In GTNetS, a simulation is cut into a possible large number of smaller simulations, each one simulating a subset of the nodes. For nodes of one subset to be able to communicate with node of another subset, related links need to be handled by two sub-simulators. These are called remote-link, the attached node are called remote-nodes. Routing often involve having routing tables at each node, but this leads to memory space growing quadratically with the number of nodes. Therefore, routing tables are built inside the simulator on an on-demand basis, using an optimization called Nix-routing. Finally, not to over-utilize disk space, logged events can be filtered, in order for the resulting log-file to be as small as possible.

Scalability of the GTNetS platform has been demonstrated in \cite{zhang05performance}. Here GTNetS was used to evaluate performance of the AODV wireless network routing protocol. Study was performed on a up to 50,000 node network. Apart from this paper, simulation were run using GTNetS on a 450 Mhz Sun UltraSPRAC-II machine running the Solaris operating system, on a 17 machine-wide Linux-based cluster platform (136 CPUs), and on a 750 server-wide supercomputer at the Pittsburgh Supercomputing Center (3000 CPUs).

Although it is not this document's primary scope, we will quickly describe how GTNetS achieves good scalability \cite{riley03large}. Simulator scalability is limited mainly by the amount of memory available (including virtual memory). Other limitations include limited disk space for log file, and simulation run-time duration limitations. GTNetS has been designed with 3 goals in mind: (1) reducing the event list size, (2) offering optimized memory management and (3) reducing the size of the log file. By reducing the number of events, simulation time is reduced. In order to do so, GTNetS uses FIFO receive queues, abstract packet queuing techniques, and timers buckets. The overall memory footprint is reduced by using NIx-vector routing and optimizing the packet representation. Finally, the log file size is reduced by allowing the user to define exactly what has to be logged. What's more, built-in statistics collection avoids having to build statistics after run-time on a huge log file.

\begin{figure}
\centering
\includegraphics[width=0.40\textwidth]{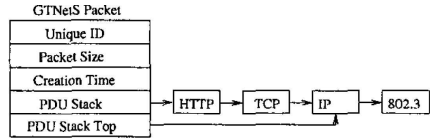}
\caption{A packet is represented efficiently in GTNetS [drawn from \cite{riley03large}].}
\label{figure:gtsnets_packet}
\end{figure}


\subsection{ease of use and extensibility}


\subsubsection{using GTNetS}

GTNetS is written entirely in object-oriented C++. The person in charge of running the simulation starts writing the main C++ program. After including the appropriate packages (especially the simulator.h header file), she or he builds the topology by instantiating the appropriate node, interface and link objects. Additional protocols, random flow generators and start/stop times and refined and the simulation is started. After successfully compiling the main program -- using any C++ compliant compiler --, it is linked with the GTNetS object libraries. The resulting executable binary is simply executed as any other application. The complete GTNetS manual can be found online \cite{rileyusing}.


\subsubsection{Implementing node mobility in GTNetS}

New application emerge for wireless ad-hoc and sensor networks. In a pervasive computing scenario, PDAs, mobile phones or even smart watches people wear could communicate and exchange information (such as propagating interesting location-specific information). In order to gather consistent performance results, simulation can be used. \cite{cavin02accuracy} has shown us the importance of the physical layer model. Here we face another challenge: the mobility model. The mostly used random waypoint model \cite{karp00gpsr} is close to pedestrian movement in an open space such as an airport hall. This is nevertheless not realistic in most scenarios, and a more general mobility model is needed.

Konishi et al. have proposed a pedestrian mobility model \cite{konishi05mobireal}. This model is best suited for urban pedestrian shopping street. Modeling starts with inputting information on the simulation field. Building, walls, and roads are drawn, and hot spots are defined (very attractive shops for example). Rules will then model the movement of pedestrian. Anti-collision algorithms model the way humans walk in crowded streets without bumping into each other. An interesting fact is that the application affects the pedestrian's movement: when a user receives information about a specially attractive hot spot, it will walk that way. A rule also makes the simulated pedestrians prefer non-crowded streets if they know where they are going.

\begin{figure}
\centering
\includegraphics[width=0.60\textwidth]{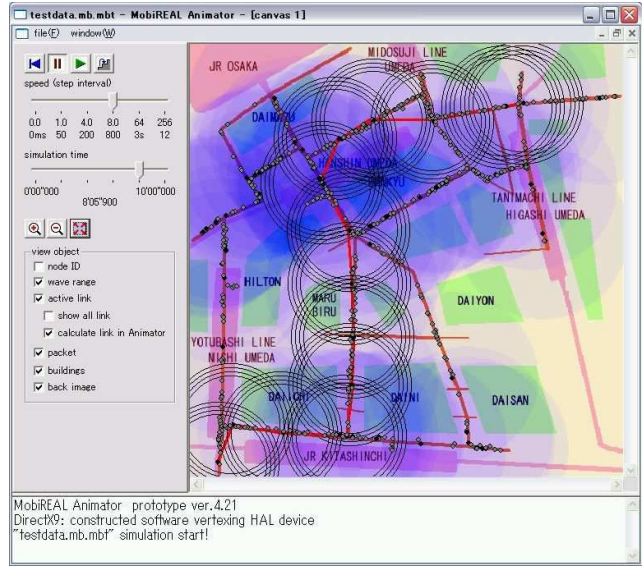}
\caption{MobiREAL's cute interface animates the model of shopping streets in downtown Osaka [drawn from \cite{konishi05mobireal}].}
\label{figure:mobireal}
\end{figure}

This mobility model is injected in the MobiREAL simulation environment, which also offers a very nice looking interface (see Figure \ref{figure:mobireal}). The underlying simulation engine is GTNetS, which provides the physical and MAC layer models.

\subsubsection{Modeling wireless sensor networks with GTSNetS}

As wireless sensor networks can potentially comprise several tens of thousands of nodes, scalability in the simulation environment is an important issues. What's more, as WSNs are very application dependent, it should be easy to add communication protocols, for example to an existing simulator. GTNetS offers the desired scalability and extensibility, so it has been extended to support WSNs. The Georgia Tech Sensor Network Simulator (GTSNetS) \cite{ouldahmedvall05simulation} offers various GTNetS C++ wireless sensor models. Apart from scalability -- which is inherited from GTNetS -- GTSNetS focuses on energy consumption.

\begin{figure}
\centering
\includegraphics[width=0.60\textwidth]{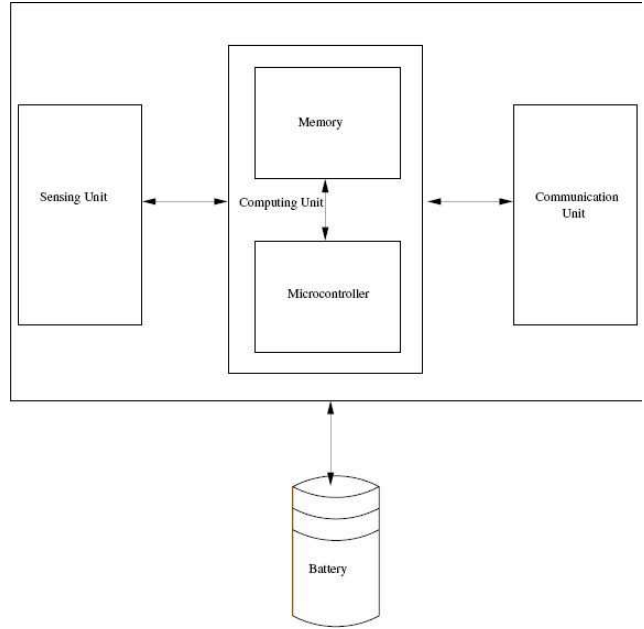}
\caption{In GTSNetS, a sensor is composed of sensing, communication, computation units, and a battery [drawn from \cite{ouldahmedvall05simulation}].}
\label{figure:gtsnets_sensor_model}
\end{figure}

Figure \ref{figure:gtsnets_sensor_model} illustrates the wireless sensor model used by Ould-Ahmed-Vall et al. Logically, each one of those component will have a tunable C++ object associated, which all include several energy models. Details of those models can be found in \cite{ouldahmedvall05simulation}, and the GTSNetS extension to GTNetS can be found online in \cite{siteGtnets}.


\section{Concluding Remarks}

\begin{table}
\begin{tabular}{|l|l|l|l|}
\hline
\textbf{Simulator} & \textbf{Cost} & \textbf{Ease of use} & \textbf{Extensibility} \\
\hline
\hline
OPNET Modeler & expensive & graphical & hard \\
\hline
ns-2 & free & flat model & possible, open-source \\
\hline
GTSNetS & free & object-oriented & easy, open-source \\
\hline
\end{tabular}
\\
\begin{tabular}{|l|l|l|l|}
\hline
\textbf{Simulator} & \textbf{Scalability} & \textbf{physical layer} & \textbf{power consump.} \\
\hline
\hline
OPNET Modeler & 1,000 nodes & 11 stage fine model & none \\
\hline
ns-2 & 1,000 nodes & 3 models & none \\
\hline
GTSNetS & 100,000+ nodes & complete 802.11 & Wireless Sensors \\
\hline
\end{tabular}
\caption{Comparing the different simulation environments.}
\label{table:comparison}
\end{table}

Table \ref{table:comparison} summarizes this document. We are interested in simulating power-aware self-organizing protocols for Wireless Sensor Networks. As WSN can be composed of a large number of nodes, scalability is a major factor in choosing a simulation platform. What's more, as the results we are interested in gathering are related to power consumption, having a wireless sensor power consumption model is very attractive.

This leads us to the conclusion that, for out needs, the best suited simulation platform is GTNetS. Future work include mastering this platform. It would be interesting to see what energy consumption models have been implemented, and to see if they are close to the models we have defined. What's more, as for scalability, it would be interesting to see how big a WSN we can simulate on a single processor personal computer, and what the overhead would be to use a Linux-based cluster.

\bibliographystyle{IEEEtran}
\bibliography{bibliography}

\begin{thebibliography}{10}
\providecommand{\url}[1]{#1}
\csname url@rmstyle\endcsname
\providecommand{\newblock}{\relax}
\providecommand{\bibinfo}[2]{#2}
\providecommand\BIBentrySTDinterwordspacing{\spaceskip=0pt\relax}
\providecommand\BIBentryALTinterwordstretchfactor{4}
\providecommand\BIBentryALTinterwordspacing{\spaceskip=\fontdimen2\font plus
\BIBentryALTinterwordstretchfactor\fontdimen3\font minus
  \fontdimen4\font\relax}
\providecommand\BIBforeignlanguage[2]{{%
\expandafter\ifx\csname l@#1\endcsname\relax
\typeout{** WARNING: IEEEtran.bst: No hyphenation pattern has been}%
\typeout{** loaded for the language `#1'. Using the pattern for}%
\typeout{** the default language instead.}%
\else
\language=\csname l@#1\endcsname
\fi
#2}}

\bibitem{fujimoto03how}
R.~M. Fujimoto, K.~S. Perumalla, A.~Park, H.~Wu, M.~H. Ammar, and G.~F. Riley,
  ``Large-scale network simulation: How big? how fast?'' in \emph{11th IEEE
  International Symposium on Modeling, Analysis, and Simulation of Computer and
  Telecommunications Systems (MASCOTS)}.\hskip 1em plus 0.5em minus 0.4em\relax
  Orlando, FL, USA: IEEE, October 2003, pp. 116--124.

\bibitem{cavin02accuracy}
D.~Cavin, Y.~Sasson, and A.~Schiper, ``On the accuracy of manet simulators,''
  in \emph{Workshop on Principles Mobile Computing (POMC)}.\hskip 1em plus
  0.5em minus 0.4em\relax Toulouse, France: ACM, October 2002, pp. 38--43.

\bibitem{haq05simulation}
F.~Haq and T.~Kunz, ``Simulation vs. emulation: Evaluating mobile ad hoc
  network routing protocols,'' in \emph{International Workshop on Wireless
  Ad-Hoc Networks (IWWAN)}, London, United Kingdom, May 2005.

\bibitem{shnayder04simulating}
V.~Shnayder, M.~Hempstead, B.~Chen, G.~W. Allen, and M.~Welsh, ``Simulating the
  power consumption of large scale sensor network applications,'' in
  \emph{Conference on Embedded Networked Systems (SenSyc)}.\hskip 1em plus
  0.5em minus 0.4em\relax Baltimore, Maryland, USA: ACM, November 2004.

\bibitem{siteOPNET}
\BIBentryALTinterwordspacing
``Opnet technologies, inc.'' last visited on 16/09/2006. [Online]. Available:
  \url{http://www.opnet.com/}
\BIBentrySTDinterwordspacing

\bibitem{siteNs2}
\BIBentryALTinterwordspacing
``Network simulator 2 home page,'' last visited on16/09/2006. [Online].
  Available: \url{http://nsnam.sourceforge.net/}
\BIBentrySTDinterwordspacing

\bibitem{siteGtnets}
\BIBentryALTinterwordspacing
``Gtnets home,'' last visited on 18/09/2006. [Online]. Available:
  \url{http://www.ece.gatech.edu/research/labs/MANIACS/GTNetS/}
\BIBentrySTDinterwordspacing

\bibitem{tanenbaum02computer}
A.~S. Tanenbaum, \emph{Computer Networks, Fourth Edition}, A.~S. Tanenbaum,
  Ed.\hskip 1em plus 0.5em minus 0.4em\relax Prentice Hall, August 2002.

\bibitem{siteGnuplot}
\BIBentryALTinterwordspacing
``gnuplot homepage,'' last visited on 18/09/2006. [Online]. Available:
  \url{http://www.gnuplot.info/}
\BIBentrySTDinterwordspacing

\bibitem{fall06ns}
\BIBentryALTinterwordspacing
K.~Fall and K.~Varadhan, \emph{The ns manual}, January 2006. [Online].
  Available: \url{http://www.isi.edu/nsnam/ns/doc/ns_doc.pdf}
\BIBentrySTDinterwordspacing

\bibitem{siteNam}
\BIBentryALTinterwordspacing
``Nam: Network animator,'' last visited on 18/09/2006. [Online]. Available:
  \url{http://www.isi.edu/nsnam/nam/}
\BIBentrySTDinterwordspacing

\bibitem{kurkowski05visualization}
S.~Kurkowski, T.~Camp, N.~Mushell, and M.~Colagrosso, ``A visualization and
  analysis tool for ns-2 wireless simulations: inspect,'' in
  \emph{International Symposium on Modeling, Analysis, and Simulation of
  Computer and Telecommunication Systems (MASCOTS)}.\hskip 1em plus 0.5em minus
  0.4em\relax Atlanta, GA, USA: IEEE, September 2005, pp. 503--506.

\bibitem{riley03large}
G.~Riley, ``Large-scale network simulations with gtnets,'' in \emph{Winter
  Simulation Conference}, vol.~1.\hskip 1em plus 0.5em minus 0.4em\relax New
  Orleans, Louisiana, USA: IEEE, December 7-10 2003, pp. 676--684.

\bibitem{riley03georgia}
G.~F. Riley, ``The georgia tech network simulator,'' in \emph{workshop on
  Models, methods and tools for reproducible network research}.\hskip 1em plus
  0.5em minus 0.4em\relax Karlsruhe, Germany: ACM SIGCOMM, 2003, pp. 5--12.

\bibitem{zhang05performance}
X.~Zhang and G.~F. Riley, ``Performance of routing in very large-scale mobile
  wireless ad hoc networks,'' in \emph{International Symposium on Modeling,
  Analysis, and Simulation of Computer and Telecommunication Systems
  (MASCOTS)}.\hskip 1em plus 0.5em minus 0.4em\relax Atlanta, GA, USA: IEEE,
  September 2005, pp. 115--122.

\bibitem{rileyusing}
\BIBentryALTinterwordspacing
G.~F. Riley, \emph{Using the Georgia Tech Network Simulator}, Georgia Institute
  of Technology, Atlanta, GA, USA. [Online]. Available:
  \url{www.ece.gatech.edu/research/ labs/MANIACS/GTNetS/doc/gtnets.pdf}
\BIBentrySTDinterwordspacing

\bibitem{karp00gpsr}
B.~Karp and H.~Kung, ``Gpsr: Greedy perimeter stateless routing for wireless
  networks,'' in \emph{Annual International Conference on Mobile Computing and
  Networking (Mobicom)}.\hskip 1em plus 0.5em minus 0.4em\relax ACM, August
  2000, pp. 243--254.

\bibitem{konishi05mobireal}
K.~Konishi, K.~Maeda, K.~Sato, A.~Yamasaki, H.~Yamaguchi, K.~Yasumoto, and
  T.~Higashino, ``Mobireal simulator - evaluating manet applications in real
  environments,'' in \emph{International Symposium on Modeling, Analysis, and
  Simulation of Computer and Telecommunication Systems (MASCOTS)}.\hskip 1em
  plus 0.5em minus 0.4em\relax Atlanta, GA, USA: IEEE, September 2005, pp.
  499--502.

\bibitem{ouldahmedvall05simulation}
E.~Ould-Ahmed-Vall, G.~F. Riley, B.~S. Heck, and D.~Reddy, ``Simulation of
  large-scale sensor networks using gtsnets,'' in \emph{International Symposium
  on Modeling, Analysis, and Simulation of Computer and Telecommunication
  Systems (MASCOTS)}.\hskip 1em plus 0.5em minus 0.4em\relax Atlanta, GA, USA:
  IEEE, September 2005, pp. 211--218.

\end{thebibliography}

\end{document}